\documentclass[12pt]{article}
\usepackage{amssymb}
\usepackage{feynmp}

\newcommand{\no}{\nonumber}
\newcommand{\eq}{\begin{equation}} 
\newcommand{\eqx}{\end{equation}}
\newcommand{\eqn}{\begin{eqnarray}} 
\newcommand{\eqnx}{\end{eqnarray}}

\newcommand{\f}[2]{\frac{#1}{#2}}
\newcommand{\lra}{\longrightarrow}
\newcommand{\cor}[1]{\left\langle{#1}\right\rangle}

\newcommand{\hyp}[4]{{ }_2F_1(#1,#2;#3|#4)}

\newcommand{\om}{\omega}
\newcommand{\dl}{\delta}
\newcommand{\gm}{\gamma}
\newcommand{\lm}{\lambda}
\newcommand{\al}{\alpha}
\newcommand{\bt}{\beta}

\newcommand{\xb}{\bar{x}}
\newcommand{\zb}{\bar{z}}
\newcommand{\qb}{\bar{q}}
\newcommand{\kb}{\bar{k}}
\newcommand{\wb}{\bar{w}}
\newcommand{\htl}{{\tilde{h}}}
\newcommand{\ft}{\tilde{f}}
\newcommand{\omt}{\tilde{\om}}

\newcommand{\CC}{\mathbb{C}}
\newcommand{\ZZ}{\mathbb{Z}}
\newcommand{\nnn}{\mathbb{N}}
\newcommand{\rrr}{\mathbb{R}}

\newcommand{\EE}{{\cal E}}
\newcommand{\GG}{{\cal G}}
\newcommand{\slc}{$SL(2,\CC)$\ }

\newcommand{\ehk}{\EE^h_{k,q}}
\newcommand{\hh}{{\cal H}}

\newcommand{\ket}[1]{\left|{#1}\right\rangle}
\newcommand{\vac}{\ket{0}}
\newcommand{\Phih}{\Phi^h(x  )}
\newcommand{\tah}{\times \{a.h.\}}

\newcommand{\arr}[4]{
\left(\begin{array}{cc}
#1&#2\\
#3&#4
\end{array}\right)
}

\newcommand{\tran}[1]{\mathfrak{T}_{#1}}
\newcommand{\sque}[1]{\mathfrak{h}_{#1}}

\newcommand{\hz}{h_0}

\newcommand{\hp}{h_p}
\newcommand{\aqqq}{A^{\hz,...,\hp}(q_0,...,q_p)}

\newcommand{\raz}{\rho_{a_0}}
\newcommand{\rao}{\rho_{a_1}}
\newcommand{\rbz}{\rho_{b_0}}
\newcommand{\rbo}{\rho_{b_1}}
\newcommand{\rpz}{\rho_{p_0}}
\newcommand{\rpo}{\rho_{p_1}}
\newcommand{\kaz}{k_{a_0}}
\newcommand{\kao}{k_{a_1}}
\newcommand{\kpz}{k_{p_0}}
\newcommand{\kpo}{k_{p_1}}
\newcommand{\Eb}{\bar{E}}

\newcommand{\rr}[4]{#1, {\it #2 \/}{\bf #3} #4}

\begin{document}

\title{Conformal invariance and QCD Pomeron vertices in the $1/N_c$ limit}
\author{R.A. Janik$^{a,b}$  and R. Peschanski$^a$ \\ \\
$^a$Service de Physique Theorique  CEA-Saclay \\ F-91191
Gif-sur-Yvette Cedex, France\\
$^b$M.Smoluchowski Institute of Physics, Jagellonian University\\ Reymonta
4, 30-059 Cracow, Poland}
\maketitle

\begin{abstract}
Using the dipole framework for QCD at small $x_{Bj}$ in the $1/N_c$ limit, we derive the expression of the $1\!\to\!p$ dipole multiplicity density in momentum space. This gives an analytical expression for the $1\!\to\!p$ QCD Pomeron amplitudes  in terms of   one-loop integration of effective vertices
in transverse  momentum.  Conformal invariance and a Hilbert space construction for dipole correlation functions are the main tools of the derivation. Relations with conformal field theories in the classical limit are discussed.
\end{abstract}

\section{Introduction}

QCD being a  scale-invariant field theory at classical level, it is natural to discuss whether scaling symmetries can be exactly or approximately preserved in some kinematical regime. For instance, in high-energy reactions, scale invariance in the transverse plane or its extension to 2-dimensional conformal invariance have  been discussed. The existence of such  symmetries could give new insights into the properties of the theory.  However, the absence of a solution of QCD at strong coupling leaves the discussion open concerning the non-perturbative aspects of the problem. The interest of  resummation of  perturbative QCD at high energy (small Bjorken $x_{Bj}$) is to allow a theoretical investigation of high-energy amplitudes in a regime which can be computed from our present knowledge of the theory.

The main outcome of the perturbative resummation calculations at leading $\log x_{Bj}$ in the relevant domain is the emergence of a Regge singularity, the so-called BFKL Pomeron, which is a compound state of 2 reggeized gluons \cite {bfkl}. In impact parameter space, the BFKL Pomeron contributions can be expressed \cite {mueller} in terms
of the formation and  interactions of colorless $q \bar q$ dipoles in the $\f 1{N_c}$ limit of small $x_{Bj}$-QCD. Interestingly enough, the calculation of various processes in both BFKL and  QCD dipole picture frameworks
exhibits  properties related to scale and conformal invariance.

First of all, it was shown \cite {Lipatov} that the BFKL equation possesses a basic 
global \slc conformal invariance which is an useful technical tool. Indeed, conformal invariance tools appear in other calculations related to the BFKL Pomerons, for instance the $2\to 4$ gluon amplitudes \cite {lotter}. Using the same invariance it is possible to derive \cite {nav1} an equivalence between s-channel dipole-dipole interactions and the t-channel BFKL Pomeron. 

These conformal invariance properties, and the possible extension to conformal field theories  (CFT) have been recently pointed out in relation with studies on multi-pomeron (multi-dipole) processes. The solution of the equations for the $1\!\to\!p$ dipole multiplicity density in impact parameter space is found to be identical to Shapiro-Virasoro closed string amplitudes \cite{pesch}. In fact, the result of reference \cite{pesch} goes beyond   \slc invariance
 since it relies on  the specific form  of the BFKL kernel. It was
proposed \cite{kor} that the interaction vertices of three BFKL Pomerons can be
related to correlation functions in two-dimensional conformal field
theory (CFT), satisfying conformal bootstrap relations. It was checked that this  leads to the same amplitudes as found in the dipole formulation.

In fact, the direct calculation from QCD Feynman diagrams is only presently available for the  $2\to 4$ gluon amplitudes \cite {lotter}. It can easily be checked that,   when considering three colour-singlet gluon-gluon (e.g. QCD Pomeron) channels in the $1/N_c$ limit, it leads to the  expression obtained in the dipole formulation. No direct calculation of amplitudes with more channels are yet available.

Our goal in the present paper is to investigate the general   $1\!\to\!p$ QCD Pomeron amplitudes using  relations between
BFKL-dipole amplitudes and correlation functions of suitably defined two-dimensional operators. The main tool of our derivations is to use a Hilbert space construction of the operators representing the dipole physical states and  their correlation functions. Our main observation is that a   formulation of the correlation functions can be given in the momentum space representation leading to a simple and physically appealing expression of the QCD Pomeron   amplitudes. This representation is realized in terms of one-loop amplitudes built from explicit  effective Pomeron vertices.

The plan of our paper is the following. In section {\bf 2} we introduce a Hilbert space construction of dipole correlation functions in impact parameter space.  In section {\bf 3} we derive the corresponding momentum space representation which, in the following part  {\bf 4}, is shown to possess a simple algebraic and geometrical formulation. In {\bf 5}, we  discuss the CFT interpretations of our results and give our conclusions in section {\bf 6}. Two appendices complete the paper.
\vspace{-2pt}

\section{Hilbert space construction of dipole correlation functions}

The $1\!\to\!p$ dipole multiplicity
density (i.e. the probability for finding $p$ dipoles $\rbz
\rbo,$...$,\rpz \rpo$ in an initial one $\raz \rao$) is obtained
\cite{pesch} from the solution of a integro-differential
equation\footnote{The multiplicity density is normalized by the unit
of initial dipole size squared $d_p \equiv {n_p} /{\rho^2_{a_0a_1}},$ where
$n_p$ is the quantity computed in \cite{pesch}.}. It reads: 
\eqn
\!\!\!\!d_p(\raz\rao,...,\rpz\rpo)=\int dh_0dh_1 ... dh_p \ \f{1}{2a_{h_0}  ... a_{h_p}}\ \times \nonumber\\
\times \f{1}{\om-\om_{h_0}}\ \f{1}{\om_{h_1}+...+\om_{h_p}-\om} \ \f{1}{|\rho_{a_0a_1} ... \rho_{p_0p_1}|^2}
\times \nonumber\\
\!\!\times \int d^2\rho_\al
d^2 \rho_\bt ... d^2\rho_\pi \ \Eb^{h_0}(\rho_{a_0\al},\rho_{a_1\al})
 ... 
\Eb^{h_p}(\rho_{0\pi},\rho_{1\pi})\ B_{1\to p},
\label{np}
\eqnx
where
$a_h=\f{\pi^4/2}{\nu^2+n^2/4}$
and
\begin{equation}
B_{1\to p}=\int \frac{d^{2}\rho _{0}...d^{2}\rho _{p}}
{\left| \rho _{01}\ \rho _{12} ... \rho _{p0}\right| ^{2}}\ E^{h_0}{\left( \rho _{0\alpha},\rho _{1\alpha} \right)}
E^{h_1}\left( \rho
_{1\bt },\rho _{2\bt }\right) ... E^{h_p}\left( \rho _{p\pi},\rho _{0\pi}\right)\ , 
\label{2}
\end{equation}
with $\rho _{ij}=\rho _{i}\!-\!\rho _{j} \left({\rm  resp.}\ \bar{\rho}
_{ij}=\bar{\rho _{i}}\!-\!\bar{\rho _{j}}\right) .$
In formula (\ref{2}),
\eq
E^{h}\left( \rho _{i\delta },\rho _{j\delta }\right) =
\left( -1\right)^{n}
\left( \frac{\rho _{ij}}{\rho _{i\delta }\rho _{j\delta }}\right)^{h}
\left( \frac{\bar{\rho} _{ij}}{\bar{\rho} _{i\delta }\bar\rho _{j\delta}}\right)^{\tilde{h}}
\label{3}
\eqx
\noindent  are \cite{Lipatov} the \slc eigenvectors
labeled by the quantum numbers of the irreducible unitary representations, namely $h =i\nu +\frac {1-n}2$, $ \tilde h
= 1\!- \!\bar h = i\nu +\frac {1+n}2,$ ($n\!\in \! {\ZZ}$, ${\nu }\!\in  \!{\rrr}$). The variable $\om$ in formula (\ref{np}) is the Mellin-transform conjugate of rapidity while the functions $\om_{h_0},...,\om_{h_p}$ are the BFKL eigenvalues associated with the $p+1$ dipoles involved in the process, i.e.
\begin{equation}
\omega_h =\frac{2\alpha N_{c}}{\pi}\Re e\left\{ \psi
\left( 1\right) -\psi \left(\frac {1+n}2 +i\nu \right) \right\}.
\label{omega}
\end{equation}

In the following, we shall consider the expression of   the functions $B_{1\to p}$  as  correlation functions, namely 
\eq
\label{e.dipcor}
B_{1\to p} \equiv \cor{0|\Phi^{h_0}(\rho_\al) \Phi^{h_1}(\rho_\bt) ...\Phi^{h_p}(\rho_\pi)|0}
\eqx
where the $\Phi^h(x)$ are   suitably defined operators. Such a
formulation has been discussed in ref. \cite{Lipatov} in comparison
with the general \slc invariant Green functions
\cite{pol}. Along these lines, an  approach \cite{kor}
considered correlation functions of  quasi-primary operators
\cite{bpz}, where 4-point correlation functions obey conformal
bootstrap 
constraints. in the following we shall derive an explicit construction of the relevant operators in a different context.

An inspiring example  for our derivation is the expression
of correlation functions  in the mini-superspace (classical)
limit \cite{seiberg,polchinsky} of the coset $H^+_3 \equiv SL(2,\CC)/SU(2)$ Wess-Zumino-Novikov-Witten (WZNW) conformal field theory \cite{Teschner}.
In this case, the derivation is based on an explicit construction of the primary fields (operators) $\Psi ^j(x),$ where $j$ labels some \slc representation and $x,\bar x$ are auxiliary variables, that is, not belonging to the 2-dimensional coordinate space on which the conformal fields are defined. In fact these auxiliary variables span the set of states in a given infinite-dimensional representation. The operators act on the Hilbert space $L^2(H^+_3)$ by multiplication by functions forming an (overcomplete) basis of the Hilbert space. The correlation functions are realized as the expectation value of a product of these operators evaluated in a state defining the  \slc invariant vacuum.

Starting with our derivation, the following ingredients are defined in order to give an  operator realization of the correlation functions (\ref{e.dipcor}): 

i) The Hilbert space is choosen to be the set of functions $f \in \hh=L^2(\CC^2)$, with a representation of  the action of the \slc group
given by  
\eq
\label{4}
[T(g)f](\rho_1,\rho_2)=(c\rho_1+d)^{-1}(c\rho_2+d)^{-1}\tah \ f(\rho'_1,\rho'_2)
\eqx
where $\rho'=\f{a\rho+b}{c\rho+d},$ and the shortened notation $\tah$ is used for multiplying by the antiholomorphic counterpart. The \slc matrix $g$ is by definition
\eq
\label{e.g}
\quad\quad g=\arr{a}{c}{b}{d} \ .
\eqx

ii) The \slc invariant vacuum vector is $\vac=\dl(\rho_1\!-\!\rho_2)$. Indeed, it is easy to check that, using (\ref{4}),  it verifies the relation $T(g)\vac=\vac.$

\medskip

iii) The operators $\Phih$ act by
\eq
\label{6}
[\Phih f](\rho,\rho'')=\int \f{d^2\rho'}{|\rho'\!-\!\rho|^2} E^h(\rho\!-\!x,\rho'\!-\!x)
f(\rho',\rho'')\ .
\eqx

To check that we indeed recover the correct expression (\ref{e.dipcor}) we see that 
acting with the chain of operators $\Phi^{h_0}(\rho_\al) \Phi^{h_1}(\rho_\bt)\ ...
\Phi^{h_p}(\rho_\pi)$ on the vacuum
vector $\vac$ gives the following function of the variables $\rho_0$ and $\rho_{p+2}$
\eq
\int \f{d^2\rho_1 ...  d^2\rho_{p+1}}{|\rho_{01}...\rho_{p(p+1)}|^2}
E^{\hz}(\rho_{0\al},\rho_{1\al}) ...
E^{\hp}(\rho_{p\pi},\rho_{(p+1)\pi})\ \dl(\rho_{p+1}\!-\!\rho_{p+2})\ .  
\eqx
Now taking the scalar product with the vacuum vector amounts to
multiplying the above expression by $\dl(\rho_0-\rho_{p+2})$ and integrating
over $\rho_0$ and $\rho_{p+2}$. We then recover the dipole formula
(\ref{2}). 

The expression (\ref{6}) enables us to  exhibit the conformal
properties of the dipole amplitudes. One shows that
\eq
T(g)\Phi^h(x') T(g)^{-1}=(cx+d)^{2h}\tah \ \Phi^h(x)
\eqx
where $x'=\f{ax+b}{cx+d}$. 

In the following we will go to momentum space and find a compact
realization, which exhibits  remarkably simple properties with respect to
\slc transformations of momenta. Also it has a direct physical
significance as giving an expression for the $1\to p$ multiplicity density of
dipoles in the momentum representation, and thus the $1\to p$ QCD Pomeron amplitudes in the $1/N_c$ limit.

\section{Correlation functions in momentum space}

The action of the operators $\Phi^h(x)$ defined in (\ref{6}) in the Fourier transformed space of functions 
\eq
\ft(k,k')=\f 1{(2\pi)^4}\int d^2\rho d^2\rho' \ e^{i(k\rho+k'\rho')} f(\rho,\rho'),
\eqx
 is given by:
\eq
[\Phih f](k,k'')=\int d^2k' \int d^2\rho d^2\rho' \ e^{ik\rho-ik'\rho'} \f{E^h(\rho\!-\!x,\rho'\!-\!x)}{|\rho-\rho'|^2}  \ft(k',k'')\ .
\eqx
A substantial simplification occurs when we will consider the Fourier
transform of the $\Phi^h$ operator with respect to $x$:
\eq
\Phi^h(q)=\int d^2x \ e^{-iqx}\Phi^h(x)\ .
\eqx
Now the vacuum becomes $\vac=\dl(k+k')$ and the action of the
operators becomes just a multiplication by a function combined with a
shift in the argument:
\eq
[\Phi^h(q) \ft](k,k')=\EE^h_{k,q} \cdot \ft(k\!-\!q,k')\ ,
\eqx
where we have defined
\eq
\ehk\equiv \int d^2\rho d^2\rho' \ e^{iq\rho'+ik(\rho\!-\!\rho')}\ 
\f{E^h(\rho,\rho')}{|\rho\!-\!\rho'|^2}\ . 
\label{Epsilon}
\eqx
Since the r\^ole of the second coordinate is  to
generate a delta function for the conservation of momentum, we may redefine
the Hilbert space, vacuum state and operators in such a way that:
\eq
\cor{\Phi^{h_0}(q_0)\ldots\Phi^{h_p}(q_p)}= \dl(q_0+\ldots+q_p) \ 
\cor{\phi^{h_0}(q_0)\ldots\phi^{h_p}(q_p)}\ ,
\label{corfourier}
\eqx
where the $\phi^h(q)$ act on the Hilbert space $L^2(\CC)$ with the
vacuum $\vac=1$ by
\eq
\label{e.constrfin}
[\phi^h(q) f](k)= \ehk \cdot f(k-q) 
\eqx

The result of the Fourier transform is
\eqn
\label{e.aform}
A^{h_0, ... ,h_p}(q_0,...,q_p)&\equiv& \cor{\phi^{h_0}(q_0) ...
\phi^{h_p}(q_p)}  \nonumber\\
&=&
\int d^2k \ \EE^{h_0}_{k,q_0} \EE^{h_1}_{k-q_0,q_1}... \EE^{h_p}_{k-q_0-...-q_{p-1},q_p}\ .
\eqnx
Thus, all dipole correlators are just expressed by a
{\em single} integral over a product of  $\ehk$ functions. 

\setlength{\unitlength}{1mm}
\begin{fmffile}{figure}

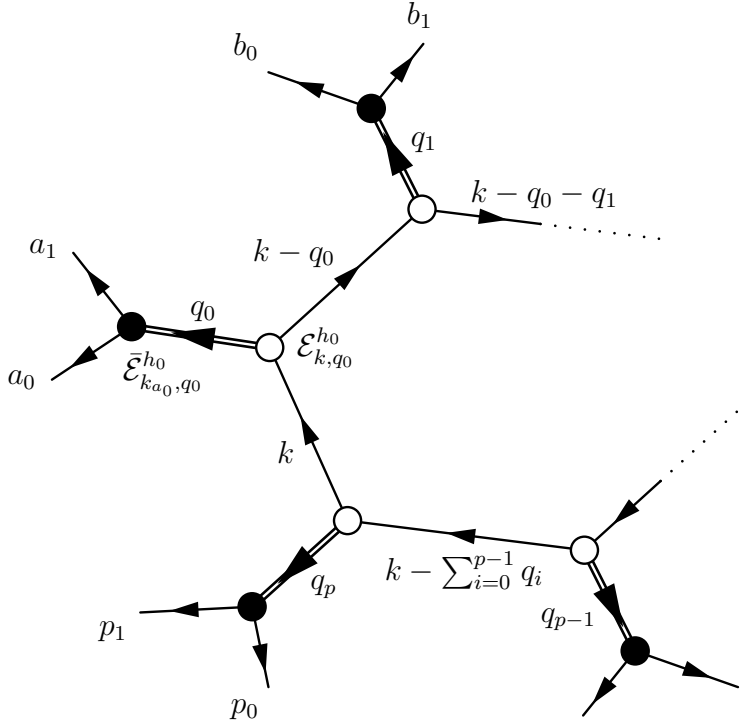
\begin{figure}[thb]
\mbox{}\hspace{1.7cm}\begin{fmfgraph*}(120,90)

\fmfcurved
\fmfsurroundn{v}{18}

\fmfv{decor.shape=circle,decor.filled=full,decor.size=0.03w}{u1,u2,u6,u5}
\fmfv{decor.shape=circle,decor.filled=empty,decor.size=0.03w}{w1,w2,w6,w5}

\fmf{fermion}{u1,v9}
\fmf{fermion}{u1,v10}
\fmf{fermion}{u2,v6}
\fmf{fermion}{u2,v7}
\fmf{phantom}{u3,v3}
\fmf{phantom}{u3,v4}
\fmf{phantom}{v1,u4}
\fmf{phantom}{v18,u4}
\fmf{fermion}{u5,v16}
\fmf{fermion}{u5,v15}
\fmf{fermion}{u6,v13}
\fmf{fermion}{u6,v12}

\fmf{heavy,la=$q_0$}{w1,u1}
\fmf{heavy,la=$q_1$}{w2,u2}
\fmf{phantom}{w3,u3}
\fmf{phantom}{w4,u4}
\fmf{heavy,la=$q_{p-1}$}{w5,u5}
\fmf{heavy,la=$q_p$}{w6,u6}

\fmf{fermion,tension=0.6,la=$k-\sum_{i=0}^{p-1}q_i$,la.side=left}{w5,w6}
\fmf{fermion,tension=0.6,la=$k$,la.side=left}{w6,w1}
\fmf{fermion,tension=0.6,la=$k-q_0$,la.side=left}{w1,w2}
\fmf{phantom,tension=0.6}{w2,w3,w4,w5}

\fmflabel{$a_0$}{v10}
\fmflabel{$a_1$}{v9}
\fmflabel{$b_0$}{v7}
\fmflabel{$b_1$}{v6}
\fmflabel{$p_0$}{v13}
\fmflabel{$p_1$}{v12}

\fmfv{la=$\EE^{h_0}_{k,,q_0}$,la.angle=0,la.dist=0.03w}{w1}

\fmffreeze

\fmfiv{la=${\bar{\EE}}^{h_0}_{k_{a_0},,q_0}$,la.angle=-75}{vloc(__u1)+(0.03w,-0.01w)}

\fmf{fermion}{w2,ww2}
\fmf{dots}{ww2,w3}
\fmf{dots}{w4,ww4}
\fmf{fermion}{ww4,w5}

\fmflabel{$k-q_0-q_1$}{ww2}

\end{fmfgraph*}
\caption{{\it Graphical representation of the  $1\to p$ QCD Pomeron amplitude. 
White
circles: internal $\ehk$ vertex functions entering the one-loop integral; Black circles: external (complex conjugate) vertices $\bar \ehk$ coupling the external gluons to the interacting BFKL Pomerons; Double lines: BFKL Pomerons.}}
\end{figure}  

\end{fmffile}

Using our Fourier transformed result (\ref{e.aform}), 
\eq
B_{1\to p} = \int d^2q_0 ... d^2 q_p \ \dl(q_0+...+q_p)\ e^{i(q_0\rho_\al +
... + q_p\rho_\pi)} \ \aqqq \ ,
\eqx
together with the one for the functions $\bar \ehk,$ (see (\ref{Epsilon})) namely
\eq
\f{\bar E^h(\rho,\rho')}{|\rho\!-\!\rho'|^2}\  
\equiv \f 1{16\pi^4}\int d^2k d^2k' \ e^{ik'\rho'+ik(\rho\!-\!\rho')}\ \bar \EE^h_{k,k'}\ ,
\label{Epsilonfourier}
\eqx
and inserting them in formula  (\ref {np}), we get
\eqn
\!\!\!\!{\cal A}(\kaz\kao,...,\kpz\kpo)\ &\equiv& \int  \f {d^2\raz}{2\pi} ...  \f {d^2\rpo}{2\pi} \ e^{-i(\raz\kaz +  ... +\rpo\kpo)} d_p(\raz...\rpo)\nonumber\\
&=& \int dh_0dh_1 ... dh_p \ \f{1}{2a_{h_0}  ... a_{h_p}}\ \times \nonumber\\
 &\times& \f{1}{\om-\om_{h_0}}\ \f{1}{\om_{h_1}+...+\om_{h_p}-\om} \  \times \GG
\ ,
\label{np1}
\eqnx
with
\eq
\GG \equiv \dl(q_0+...+q_p)\ \aqqq \ \bar {\cal E}^{h_0}_{\kaz,q_0} ... \bar {\cal E}^{h_p}_{\kpz,q_p},
\label{splitting}
\eqx
where one has to identify $q_0 = \kaz\!+\!\kao,$...$,q_p = \kpz\!+\!\kpo.$  ${\cal A}$ is  obtained as the Fourier transform in 2-dimensional momentum space of the  dipole multiplicity density $d_p$ given in formula (\ref{np}). In fact,   since dipole coordinates correspond to the Fourier transforms of the gluon momenta, ${\cal A}$ is to be interpreted as the BFKL $2\to 2p$ gluon amplitudes with  gluon-gluon singlet channels  in the $1/N_c$ limit. Thus the reduced amplitude $\aqqq$ is the $1\to p$ QCD Pomeron vertex.

Using  the expression (\ref{e.aform}) for $\aqqq,$  formulae (\ref {np1},\ref {splitting}) have a rather simple and attractive representation (see Fig.1) in terms of a one-loop integral in momentum space with vertices defined by the functions $\ehk.$ Each dipole is represented in momentum space and interacts {\it via} a 3-vertex defined by a function $\ehk$ or $\bar \ehk$ depending whether it is 
created or annihilated. The  momentum is conserved at each vertex.

In the next section we will calculate the explicit form of
the $\ehk$ vertex functions, which are the  building blocks of the obtained
expressions, and we will explore  their properties.

\section{The vertex  functions $\ehk$}

$\ehk$ is given by the double 2-dimensional Fourier integral (\ref{Epsilon}). By an appropriate sequence of 
 changes of variables, see the appendix {\bf A1}, the expression  factorizes  to
yield 
\eq
 \left[\int d^2w \ e^{i Re(w)} w^{-h}\wb^{-\htl} \right]
 \cdot \  \qb^{h\!-\!1}q^{\htl\!-\!1} \cdot I_h\left(\f{\qb}{\kb}\right)
\label{Ihk}
\eqx
where
\eq
I_h(x)=x^{1-h} \bar x^{1-\htl}\int d^2v v^{-h} (1\!-\!v)^{-h}
(1\!-\!xv)^{h-1} \tah
\label{integral}
\eqx
and the normalization is
\eq
\left[\int d^2w \ e^{i Re(w)} w^{-h}\wb^{-\htl} \right]=\f {\pi i^{h-\htl}2^{2-h-\htl}}{\gm(h)}\ .
\label {norm}
\eqx
Here we use the notation $\tah$ for both  complex conjugation and $h \to \tilde h$ while $\gm(h)=\Gamma(h)/ \Gamma(1\!-\!\htl)$.

The integral in (\ref {integral}) can be performed \cite{Navelet}  using the formula
quoted in appendix {\bf A1}:
\eq
I_h(x) = \pi\left[ \gm(1\!-\!2h)\gamma^2(h)\ 
{\cal F}_h(x){\cal F}_\htl(\xb)+ \left\{h \to 1-h\right\}
\right]
\label{appendix}
\eqx
with
\eq
{\cal F}_h(x)\equiv x^h \hyp{h}{h}{2h}{x}\ ,
\eqx
where $_2F_1$ is the standard hypergeometric function.
The final expression for $\ehk$ is therefore
\eq
\ehk=\f {\pi i^{h-\htl}2^{2-h-\htl}}{\gm(h)}\cdot \qb^{h-1}q^{\htl-1} \cdot I_h\left(\f{\qb}{\kb}\right).
\label {final}
\eqx

Using the resulting formulae (\ref {appendix}-\ref{final}), one obtains an explicit 
analytical expression for  the QCD Pomeron amplitudes. Since the vertex functions appear as their building blocks,  
let us quote a few mathematical properties of the  $\ehk.$

\bigskip
{\bf i)} $\ehk$ is an \slc matrix element:
\bigskip

Using formula (\ref{e.matel}) of the appendix {\bf A2}, one obtains a correspondance with the formula (\ref{final}) by identifying $-1/(bc)$ with $\qb/\kb$. A simple factorized
decomposition of the related group element $g$ can be found:
\eq
\label{e.gfact}
g=\underbrace{\arr{1}{0}{\kb}{1}}_{\displaystyle \tran{\kb}} \cdot
\underbrace{\arr{1}{-\qb^{-1}}{0}{1}}_{\displaystyle \sque{\qb}} 
\eqx
Here the first piece $\tran{\kb}$ is just a translation ({\em in
momentum space}) by $\kb$, while the second element $\sque{\qb}$ is the \slc transformation  $(0,\infty)\to (0,-\bar q).$ 

\bigskip
{\bf ii)} Geometrical picture:
\bigskip

Another interesting formula\footnote { The formula  has been originally obtained as a solution for  the dipole-dipole Green's function  in the impact parameter space \cite {li2}.}
 is the following\eq
\label{e.green}
I_h(x)=\int d^2k_\delta \ \bar{E}^h(k_{i\delta},k_{j\delta})
E^h(k_{i'\delta},k_{j'\delta})
\eqx
where $x=\f{k_{ij}k_{i'j'}}{k_{ii'}k_{jj'}}$ is the anharmonic ratio. Note that one uses the same \slc Eigenvectors $E^h$ as in (\ref{3}), but corresponding to \slc transformations in momentum space.
Let us now set  $k_i=0$, $k_j=\infty$, $k_{i'}=\kb$,
$k_{j'}=\kb-\qb$. We again recover  $I_h\left(\f{\qb}{\kb}\right)$ as in (\ref{Ihk}). In fact the \slc element $g$ appearing in
(\ref{e.gfact}) has a simple interpretation as an \slc element
which transforms the momenta $(0,\infty)$ into
$(\bar k,\bar k\!-\!\bar q)$. The origin of the above mentioned factorization
(\ref{e.gfact}) is now quite clear in terms of successive geometric transformations in momentum space:
\eq
(0,\infty) \lra (0,-\bar q) \lra (\bar k,\bar k\!-\!\bar q)
\eqx
By extension, the sequence of group elements corresponding to the product\\
$\EE^{h_0}_{k,q_0} ...\EE^{h_p}_{k-q_0-...-q_{p-1},q_p},$ see (\ref{e.aform}),
corresponds to \slc group transformations acting in momentum space,
and transforming the line $(0,\infty)$ into the $p+1$ lines (forming a polygon) related to  the loop of momenta shown in Fig.1. The integration over $k$ (the translation group in momentum space)
corresponds to moving the polygon all over the complex plane thus gives an intrinsic geometrical character to the QCD Pomeron correlation functions (\ref{e.aform}) and amplitudes (\ref{np1}) in momentum space. 

\section{Similarities with conformal field theories}

In the previous sections, we have constructed a solution of the $1\!\to\!p$ QCD Pomeron amplitudes as correlation functions of suitably defined operators acting in  a specific Hilbert space. Since our derivation has been  motivated by  conformal field theory constructions, it is thus natural to discuss the possible connections with conformal field theories

Let us first discuss the conjecture of \cite {kor} that dipole correlation functions  
could be interpreted as correlation functions of quasi-primary operators \cite {bpz} in some
conformal field theory.
For instance, in \cite{kor} it is
found that the  4-point correlation function, which happens to be identical to the  $1\!\to\!p, p=3$ expression (\ref {2}),   can be decomposed
into a sum of conformal blocks reminiscent of CFT:
\eq
B_{1\to 3}\equiv\cor{0|\Phi^{h_1}(z_\al) \Phi^{h_2}(z_\bt) \Phi^{h_3}(z_\gm)
\Phi^{h_4}(z_\dl)|0} =\sum_{h} C_{h_1h_2h}C_{hh_3h_4} \ |{\cal F}(x)|^2,
\eqx
where $x=\f {\rho_{\alpha \beta}\rho_{\gm \dl}}{\rho_{\alpha \gm}\rho_{\bt \dl}},$ $C$ are structure constants and 
\eq
\label{e.korconf}
{\cal F}(x)=x^{h-h_3-h_4} \hyp{h_3-h_4+h}{h_2-h_1+h}{2h}{x}\ .
\eqx
These expressions obey  CFT crossing relations \cite{kor}.

However, it turns out that the relation (\ref{e.korconf}) gives stringent constraints on the possible CFT interpretations. In particular, they restrict these CFT to be  considered in the classical limit. Indeed,  
in general CFT the form of the conformal block follows from conformal
invariance and depends only on $h_1,\ldots, h_4,h$ and the central
charge $c$. A general closed form expression is unknown but the power
series coefficients of ${\cal F}^{CFT}(x)=x^{h-h_3-h_4} (1+{\cal F}^{CFT}_1
x+{\cal F}^{CFT}_2 x^2+\ldots)$ can be calculated order by order \cite{bpz,DIF}.
Setting for example $h_1=h_2$ and $h_3=h_4$, one has \cite{DIF}
\eqn
&&{\cal F}^{CFT}_2 = \f {h(h+1)h(h+1)}{4h(2h+1)}+   \nonumber \\
&&+2\left(h_1+\f {h(h\!-\!1)}{2(2h\!-\!1)}\right)^2 \left(h_3+\f{h(h\!-\!1)}{2(2h\!-\!1)}\right)^2
\left(c+\f{2h(8h\!-\!5)}{2h+1}\right)^{-1}
\eqnx
which is similar to  the corresponding coefficient ${\cal F}_2$ of (\ref{e.korconf}) only if $c=\infty,$ that is in the classical limit. If not in this limit,  
an explicit dependence on $h_1$ and $h_3$ remains. 

Another constraint on the possible theories is provided by the overcompleteness relation \cite {Lipatov} between \slc Eigenvectors:
\eq
\label{e.refl}
E^{1-h}(\rho_{10},\rho_{20})\propto \int d^2\rho_0' \ \rho^{2h-2}_{00'}\bar\rho^{2\htl-2}_{00'} \    E^h(\rho_{10'},\rho_{20'}),
\eqx
which would mean in terms of operators that fields labeled by $h$ and $1-h$ can mix. This would be forbidden by conformal invariance, where  a
linear relation between fields with different conformal weights cannot exist, since vacuum expectation values  of a product of operators with different dimension would be non-zero. 

This problem can be circumvented by considering WZNW models \cite {DIF} with
\slc current algebra at some level $k.$ 
Then one could interpret  $h$  as
labelling representations of the current algebra and not the conformal
dimensions. In
this case fields labelled by $h$ and $1-h$ have the same conformal
dimension. In particular, in the classical limit\footnote{We do not specify here the explicit form of
this model, e.g. the possible coset structure.}
 we are led to consider $k=\infty,$ all fields have dimension $0$ and all dependence on the ``world-sheet'' coordinates vanishes. The coordinates of the fields in the correlation functions are now to be  interpreted  as auxillary
variables \cite{aux} labelling states within representations of \slc.
 
Finally, it is interesting to compare our results with the properties of the 
$SL(2,\CC)/SU(2)$-WZNW conformal field theory \cite{Teschner}. It is to be noticed that the functional
form of the conformal blocks with auxillary
variables \cite{Teschner} become exactly equal to (\ref{e.korconf}),
as expected from  the minisuperspace limit.  Let us now compare  the
action of primary operators in our case with
Ref.\cite{Teschner}. While the dipole operators act by an integral
form (\ref {6}) 
in the impact-parameter space, they act by multiplication by the
vertex function $\ehk$ in momentum space (\ref {e.constrfin}), in a
similar way as for $SL(2,\CC)/SU(2).$ However, in formula  (\ref
{e.constrfin}), there is in addition a shift in momentum in the
argument of the test function. We know that this shift is essential to
obey momentum conservation at each vertex. It will thus be important
to identify which conformal field theory structure it would correspond
to.  

In any case, a possible CFT interpretation of the dipole correlation functions 
should be compatible with the structure of an effective field theory
at one-loop level revealed by our analysis.

\label{s.cft}

\section{Conclusions}
Let us summarize the  results of our study:

i) Using a Hilbert space construction, we were able to formulate the $1\!\to\!p$ dipole multiplicity densities as correlation functions of  operators. 

ii) Going from impact parameter to momentum space, the correlation functions can be expressed as 1-loop integrals involving universal momentum dependent vertex functions $\ehk.$

iii) The resulting amplitudes correspond to the $1\to p$ QCD Pomeron vertices 
obtained in the $1/N_c$ limit through the dipole formulation.

iv) The vertex functions are explicitly given in analytic form and possess interesting algebraic and geometrical properties under global conformal \slc transformations.

These properties are very suggestive of an underlying effective 2 dimensional field theory in the transverse momentum space. We analyzed the conditions that need to be fulfilled in order to have a realization in terms of a local conformal field theory. They imply a minisuperspace (classical) limit and possibly a WZNW
symmetry structure. 

An important check of these properties would be obtained by a direct calculation of the $2 \to 2p$ gluon amplitudes. Projecting on singlet gluon-gluon channels and taking the $1/N_c$ limit would lead to a direct comparison with our formulae  (\ref {e.aform}) and (\ref {splitting}). This would provide a check of a generalized dipole-BFKL correspondence.

Further consequences of these properties can be discussed both from theoretical and phenomenological points of view.
On the theoretical ground, the symmetrical formulation  of the
correlations functions (\ref{e.dipcor},\ref{e.aform}) with respect to
the initial and final dipole fields 
seems to lead to a relation (at least for $N_c \to \infty$) between
$1\!\to\!p+r$  and $p+1\!\to\!r$ dipole multiplicity densities. This
intringuing conjecture would deserve further study for instance in the
case where 
$p+1\!=\!r$ which would correspond to a subset of the $2r\!\to\!2r$
gluon amplitudes described in the BKP framework \cite
{bet}.

On the phenomenological ground, the simple formulation of dipole splitting in terms of the $\ehk$ vertices could be useful to describe the evolution of the dipoles through the corresponding amplitudes
in momentum space. In particular, it could give a QCD-theoretical hint
for the formulation of the formation and splitting of the ``QCD string'' which is the basis of  phenomenological models of strong interactions at long distance. 

\subsection*{Acknowledgements}

We thank G.Korchemsky, S.Munier, H.Navelet, J.Wosiek and J.B.Zuber for useful discussion and remarks. One of us (R.J.)
was supported in part by KBN grant No 2P03B08614. 

\eject

\appendix
\section{$\!\!\!\!\!$1: Calculation of the vertex $\ehk$}
\eq
\ehk\equiv \int d^2\rho d^2\rho' \ e^{iq\rho'+ik(\rho\!-\!\rho')}\ 
\f 1{|\rho\!-\!\rho'|^2}\ \left(\f  {\rho\!-\!\rho'}{\rho \rho'}\right)^h\tah \ .
\eqx
Changing successively $\rho' \to u\rho$ then $w=\rho \left(\bar k +u(\qb-\kb)\right),$ one obtains a factorized form
\eq
 \left[\int d^2w \ e^{i Re(w)} w^{-h}\wb^{-\htl} \right]
 \cdot \  \kb^{h\!-\!1}k^{\htl\!-\!1}  \int d^2u u^{-h} (1\!-\!u)^{h-1}
(1\!+\!(x\!-\!1)u)^{h} \tah \ ,
\eqx
where $x=\f{\qb}{\kb}.$
Finally the change of variable $v=\f u{u-1}$ in the second integral leads to the formula (\ref{integral}) for $I_h(x).$

This integral is known in the literature \cite{Navelet}. Consider the following more general form:
\eq
{\cal I}(a_0,a_1,b_1,z)\equiv \int d^2v \ v^{a_1-1} (1-v)^{b_1-a_1-1}
(1-vz)^{-a_0} \tah 
\eqx
The result is \cite{Navelet}:
\eqn
{\cal I}(a_0,a_1,b_1,z)= \pi\biggl[\f{\gm(a_1)\gm(b_1\!-\!a_1)}{\gm(b_1)}  
\hyp{a_0}{a_1}{b_1}{z}\tah+ \nonumber\\
+\f{\gm(b_1\!-\!1)\gm(1\!-\!a_0)}{\gm(b_1\!-\!a_0)}
z^{1\!-\!b_1} \hyp{a_0\!-\!b_1\!+\!1}{a_1\!-\!b_1\!+\!1}{2\!-\!b_1}{z} \tah \biggr]\ ,
\eqnx
where  we use the notation $\tah$ for both  complex conjugation and $h \to \tilde h.$
\appendix
\section{$\!\!\!\!\!$2: \slc matrix elements}

Consider the representation of \slc given by
\eq
[T(g)f](z)=(cz+d)^{-2h}\tah \cdot f\left(\f{az+b}{cz+d}\right)
\eqx
 where
\eq
\quad\quad g=\arr{a}{c}{b}{d} \ ,
\eqx
and the set of orthogonal basis functions
\eq
\ket{\om,\omt}=z^{-h-\om}\zb^{-\htl-\omt}
\eqx
with $\om=m/2-i\sigma$, $\omt=-m/2-i\sigma$ where $m\in \nnn$ and $\sigma
\in \rrr$. These functions are
eigenfunctions of the (complex) dilatation transformations:
\eq
T\left( \arr{\lm}{0}{0}{\lm^{-1}}\right) \ket{\om,\omt}=
\lm^{-2\om}\bar{\lm}^{-2\omt} \ket{\om,\omt} 
\eqx
In particular the only dilatationally invariant state is
$\ket{0,0}$. 
The matrix element of $T(g)$ in this basis can be evaluated to yield
\eqn
\cor{\om_L,\omt_L|T(g)|\om_R,\omt_R}= a^{\omt_L^*-\om_R}
(-b)^{h-1-\omt^*_L} (-c)^{h-1+\om_R} \tah \cdot\no\\
\cdot \int d^2v \ v^{-h-\om_R}
(1-v)^{-h+\om_R} (1+\f v{bc})^{h-1-\omt^*_L}\tah \ .
\eqnx
For $\om_L=\omt_L=\om_R=\omt_R=0$ we get exactly 
\eq
\label{e.matel}
\cor{0,0|T(g)|0,0}=I_h\left(\f{-1}{bc}\right)  \ .
\eqx

\end{document}